\documentclass[twocolumn,twoside,slac_two]{revtex4}
\usepackage{graphicx}
\usepackage{fancyhdr}
\begin{document}

\title{Search for neutrino emission of gamma-ray flaring blazars with the ANTARES telescope}


\author{Dornic D.$^{1,2}$ on behalf the ANTARES Collaboration}
\affiliation{$^1$ IFIC - Instituto de Fisica Corpuscular, Edificios Investigaci´on de Paterna, CSIC - Universitat de Valencia, Apdo. de Correos 22085, 46071 Valencia, Spain}
\affiliation{$^2$ CPPM, Aix-Marseille Universit\'ee, CNRS/IN2P3, Marseille, France}

\begin{abstract}
The ANTARES telescope is well suited to detect neutrinos produced in astrophysical
transient sources as it can observe a full hemisphere of the sky at all times with a
high duty cycle. The background and point source sensitivity can be drastically
reduced by selecting a narrow time window around the assumed neutrino production
period. Radio-loud active galactic nuclei with their jets pointing almost directly towards
the observer, the so-called blazars, are particularly attractive potential neutrino point
sources, since they are among the most likely sources of the observed ultra high energy
cosmic rays and therefore, neutrinos and gamma-rays may be produced in hadronic
interactions with the surrounding medium. The gamma-ray light curves of blazars
measured by the LAT instrument on-board the Fermi satellite reveal important time
variability information. A strong correlation between the gamma-ray and the neutrino
fluxes is expected in this scenario.

An unbinned method based on the minimization of a likelihood ratio was applied to a
subsample data collected in 2008 (61 days live time). By looking for neutrinos
detected in the high state periods of the AGN light curve, the sensitivity to these sources
has been improved by about a factor 2 with respect to a standard time-integrated point
source search. First results on the search for ten bright and variable Fermi
sources are presented.

\end{abstract}

\maketitle

\thispagestyle{fancy}

\section{Introduction}

Neutrinos are unique messengers to study the high-energy universe as they are neutral and 
stable, interact weakly and therefore travel directly from their point of creation to the Earth without absorption. 
Neutrinos could play an important role in understanding the mechanisms of cosmic ray acceleration and their 
detection from a cosmic source would be a direct evidence of the presence of hadronic acceleration. 
The production of high-energy neutrinos has been proposed for several kinds
of astrophysical sources, such as active galactic nuclei (AGN), gamma-ray bursters (GRB), supernova
remnants and microquasars, in which the acceleration of hadrons may occur (see Ref.~\cite{bib:Becker} for a review). 

Radio-loud active galactic nuclei with their jets pointing almost directly towards
the observer, the so-called blazars, are particularly attractive potential neutrino point
sources, since they are among the most likely sources of the observed ultra high energy
cosmic rays and therefore, neutrinos and gamma-rays may be produced in hadronic
interactions with the surrounding medium. The gamma-ray light curves of blazars
measured by the LAT instrument on-board the Fermi satellite reveal important time
variability information on timescale of hours to several weeks, with intensities
always several times larger than the typical flux of the source in its quiescent state~\cite{bib:FermiLATAGNvariability}. 
A strong correlation between the gamma-ray and the neutrino fluxes is expected in this scenario. 

This paper presents the results of the first time-dependent search for cosmic neutrino sources by the ANTARES telescope. The data sample used in
this analysis and the comparison to Monte Carlo simulations are described in Section 2,
together with a discussion on the systematic uncertainties. The point source search
algorithm used in this time-dependent analysis is explained in Section 3. The search results are presented in
Section 4 for ten selected candidate sources.

\section{ANTARES}

The ANTARES Collaboration completed the construction of a neutrino
telescope in the Mediterranean Sea with the connection of its twelfth detector line
in May 2008~\cite{bib:Antares}. The telescope is located 40 km off the Southern coast of France
(42$^{\circ}$48'N, 6$^{\circ}$10'E) at a depth of 2475 m. It comprises a three-dimensional array of
photomultipliers housed in glass spheres (optical modules~\cite{bib:OM}), distributed along twelve
slender lines anchored at the sea bottom and kept taut by a buoy at the top. 
Each line is composed of 25 storeys of triplets of optical modules (OMs), each housing one 10-inch photomultiplier. 
The lines are subject to the sea currents and can change shape and orientation. A positioning system 
based on hydrophones, compasses and tiltmeters is used to monitor the detector geometry with an accuracy of $~10$~cm. 
The main goal of the experiment is to search for high energy neutrinos with energies greater than 100~GeV by detecting muons produced 
by the neutrino charged current interaction in the vicinity 
of the detector. Due to the large background from downgoing atmospheric muons, the telescope is optimized 
for the detection of upgoing muons as only they can originate from neutrinos.

Muons induce the emission of Cherenkov light in the sea water. The arrival time and intensity of the Cherenkov light on the OMs are digitized into hits and transmitted to shore. 
Events containing muons are selected from the continuous deep sea optical backgrounds due to natural radioactivity and bioluminescence. A detailed description of the detector 
and the data acquisition is given in~\cite{bib:Antares,bib:antaresdaq}.The arrival times of the hits are calibrated as described in~\cite{bib:TimeCalib}. A L1 hit is defined either as a high-charge hit, or as hits separated by less 
than 20~ns on OMs of the same storey. At least five L1 hits are required throughout the detector within a time window of 2.2~$\mu$s, 
with the relative photon arrival times being compatible with the light coming from a relativistic particle. Independently, 
events which have L1 hits on two sets of adjacent or next-to-adjacent floors are also selected. 

The data used in this analysis were taken in the period from September 6 to
December 31, 2008 (54720 to 54831 modified Julian days, MJD) with the twelve line detector. This period overlaps with the availability of the first data from the LAT instrument 
onboard the Fermi satellite. The corresponding effective live time is 60.8 days.
From the timing and position information of the hits, muon tracks are reconstructed
using a multi-stage fitting procedure, based on Ref.~\cite{bib:AAfit}. The initial fitting stages
provide the hit selection and starting point for the final fit. The final stage consists of a
maximum likelihood fit of the observed hit times and includes the contribution of optical
background hits. Upgoing tracks are also required to have a good reconstruction quality. The latter is quantified by a parameter, $\Lambda$ which is based on the
value of the likelihood function obtained for the fitted muon (see Ref.~\cite{bib:AAfit} for details).The angular uncertainty obtained from the muon track fit is required to be smaller than 1 degree. 
For this analysis, events are selected with $\Lambda>-5.4$. This value results in an optimal compromise between the atmospheric neutrino and muon background reduction and 
the efficiency of the cosmic neutrino signal with an assumed spectrum proportional to $E_{\nu}^{-2}$, 
where $E_{\nu}$ is the neutrino energy, which gives the best 5$\sigma$ discovery potential. The resulting sample consists of 628 events obtained in 60.8 days. 
The simulations indicate that the selected sample contains 60$\%$ atmospheric neutrinos; the rest being mis-reconstructed atmospheric muons.

\begin{figure}[ht!]
\centering
\includegraphics[width=0.4\textwidth]{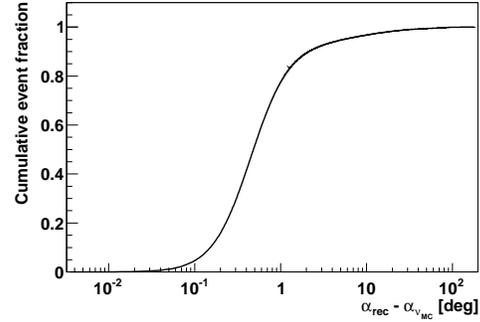}
\caption{Cumulative plot of the distribution of the angle between the true Monte Carlo neutrino direction and the reconstructed
muon direction for E$^{-2}$ upgoing neutrino events selected for this analysis.}
\label{fig:Angres}
\end{figure}

The angular resolution of the reconstructed neutrino direction can not be determined directly from the data and has to be estimated from simulation. However,
 comparison of data and Monte Carlo in which the time accuracy of the hits was degraded by up to 3~ns constrains the 
uncertainty of the angular resolution to about 0.1$^{\circ}$~\cite{bib:AAfitps}. Figure~\ref{fig:Angres} shows the cumulative distribution of the 
angular difference between the reconstructed muon direction and the neutrino direction for an assumed spectrum proportional to $E_{\nu}^{-2}$. For the 
considered period, the median resolution is estimated to be 0.5 $\pm$ 0.1 degrees.

\section{Time-dependent search algorithm}
This time dependant point source analysis is performed using an unbinned method based on a
likelihood ratio maximization. The data is parameterized as a 
mixture of signal and background. The goal is to determine, at a given point in the sky and at a given
time, the relative contribution of each component and to calculate the probability to have a signal
above a given background model. The likelihood ratio $\lambda$ is the ratio of the probability density for the hypothesis 
of background and signal ($H_{sig+bkg}$) over the probability density of only background ($H_{bkg}$):
\begin{equation}\label{eq:EQ_likelihood}
\lambda=\sum_{i=1}^{N} log\frac{\frac{n_{sig}}{N}P_{sig}(\alpha_{i},t_{i}) + (1-\frac{n_{sig}}{N})P_{bkg}(\delta_{i},t_{i})}{P_{bkg}(\alpha_{i},t_{i})}
\end{equation}

where $n_{sig}$ is the unknown number of signal events determined by the fit and N is the total number of events in the considered data sample. 
$P_{sig}(\alpha_{i},t_{i})$ and  $P_{bkg}(\delta_{i},t_{i})$ are the probability density functions (PDF) for signal and background respectively. 
For a given event \textit{i}, $t_{i}$, $\delta_{i}$ and $\alpha_{i}$ represent the time of the event, its declination and the angular separation 
from the source under consideration. 

\begin{figure}[ht!]
\centering
\includegraphics[width=0.4\textwidth]{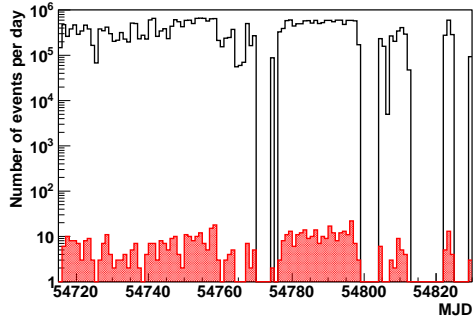}
\caption{Time distribution of the reconstructed events. Upper histogram (black line): distribution of all reconstructed events. Bottom filled histogram (red): distribution of selected upgoing events.
}
\label{fig:TimeDistri}
\end{figure}

The probability densities $P_{sig}$ and $P_{bkg}$ are factorized into a purely directional and a purely time-related component. 
The shape of the time PDF for the signal event is extracted directly from the gamma-ray light curve assuming 
proportionality between the gamma-ray and the neutrino fluxes. For signal events, the directional PDF is described by the one 
dimensional point spread function (PSF), which is the probability density of reconstructing an event at an angular distance $\alpha$ from the true source position. 
The directional and time PDF for the background are derived from the data using the observed declination distribution of the selected events 
and the observed one-day binned time distribution of all the reconstructed muons respectively. 
Figure~\ref{fig:TimeDistri} shows the time distribution of all the reconstructed events and the selected upgoing events for this analysis. Once normalized to an integral 
equal to 1, the distribution for all reconstructed events is used directly as the time PDF for the background.
Empty bins in the histograms correspond to periods with no data taking (i.e. detector in maintenance) or with very poor quality data (high bioluminescence or 
bad calibration).

\begin{figure}[ht!]
\centering
\includegraphics[width=0.4\textwidth]{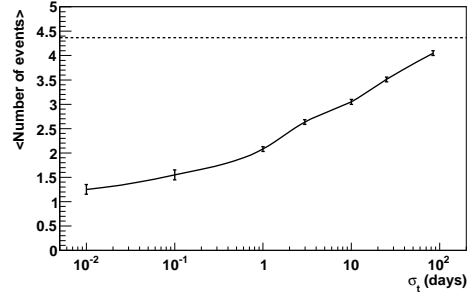}
\caption{Average number of events (solid line) required for a 5$\sigma$ discovery (50~$\%$ probability) from a single source located at a
declination of -40$^{\circ}$ as a function of the width of the flare period ($\sigma_{t}$) for the 60.8 day analysis. These numbers are compared to that obtained without
using the timing information (dashed line).}
\label{fig:Nev5sigma}
\end{figure}

The null hypothesis is given with $n_{sig}=0$. The obtained value of $\lambda_{data}$ on the data is then 
compared to the distribution of $\lambda$ given the null hypothesis. Large values of $\lambda_{data}$ compared to the distribution of $\lambda$ for the 
background only reject the null hypothesis with a confident level equal to the fraction of the scrambled trials above $\lambda_{data}$. This fraction of 
trials above $\lambda_{data}$ is referred as the p-value. The discovery potential is then defined as the average number of signal events required to 
achieve a p-value lower than 5$\sigma$ in 50~$\%$ of trials. Figure~\ref{fig:Nev5sigma} shows the average number of events required for a 5$\sigma$ discovery (50~\% C.L.) produced in one source located at a declination of -40$^{o}$ as 
a function of the total width of the flare periods. These numbers are compared to that obtained without using the timing information. Using the timing information yields 
to an improvement of the discovery potential by about a factor 2-3 with respect to a standard time-integrated point source search~\cite{bib:AAfit}.

\section{Search for neutrino emission from gamma-ray flare}

This time-dependent analysis has been applied to bright and variable Fermi blazar sources reported in the first year Fermi LAT catalogue~\cite{bib:Fermicatalogue} and in the LBAS catalogue 
(LAT Bright AGN sample~\cite{bib:FermicatalogueAGN}). The sources located in the visible part of the sky by Antares from which the averaged 1 day-binned flux 
in the high state is greater than 80 10$^{-8}$ photons.cm$^{-2}$.s$^{-1}$ above 100~MeV in the studied time period and with a significant time variability are selected. 
This list includes six flat spectrum radio quasars (FSRQ) and four BLlacs. Table~\ref{tab:Sources} lists the characteristics of the ten selected sources.

\begin{table}[ht!]
\begin{center}
\begin{tabular}{|c|c|c|c|c|c|c|}
\hline
Name & {OFGL name} & Class & {RA [$^{o}$]} & {Dec [$^{o}$]} & Redshift \\
\hline
\hline
{PKS0208-512} & {J0210.8-5100} & FSRQ & 32.70 & -51.2 & 1.003 \\
\hline
{AO0235+164} & {J0238.6+1636} & BLLac & 39.65 & 16.61 & 0.940 \\
\hline
{PKS0454-234} & {J0457.1-2325} & FSRQ & 74.28 & -23.43 & 1.003 \\
\hline
{OJ287} & {J0855.4+2009} & BLLac & 133.85 & 20.09 & 0.306 \\
\hline
{WComae} & {J1221.7+28.14} & BLLAc & 185.43 & 28.14 & 0.102 \\
\hline
{3C273} & {J1229.1+0202} & FSRQ & 187.28 & 2.05 & 0.158 \\
\hline
{3C279} & {J1256.1-0548} & FSRQ & 194.03 & -5.8 & 0.536 \\
\hline
{PKS1510-089} & {J1512.7-0905} & FSRQ & 228.18 & -9.09 & 0.36 \\
\hline
{3C454.3} & {J2254.0+1609} & FSRQ & 343.50 & 16.15 & 0.859 \\
\hline
{PKS2155-304} & {J2158.8-3014} & BLLac & 329.70 & -30.24 & 0.116 \\
\hline
\end{tabular}
\caption{List of bright variable Fermi blazars selected for this analysis~\cite{bib:FermicatalogueAGN}.}
\label{tab:Sources}
\end{center}
\end{table}

The light curves published on the Fermi web page for the monitored sources~\cite{bib:Fermimonitored} are used for this analysis. They correspond to 
the one-day binned time evolution of the average gamma-ray flux above a threshold of 100~MeV since August 2008. The high state periods are defined 
using a simple and robust method based on three main steps. Firstly, the baseline is determined with an iterative linear fit. After each fit, bins more than two sigma ($\sigma_{BL}$) 
above the baseline (BL) are removed. Secondly, seeds for the high state periods are identified by searching for bins significantly above the baseline according to the criteria: 
$(F - \sigma_{F}) > (BL + 2*\sigma_{BL}) + F > (BL + 3*\sigma_{BL})$ where F and $\sigma_{F}$ represent the flux and the uncertainty on this flux for each bin, respectively. For each seed, the adjacent bins for which the emission is 
compatible with the flare are added if they satisfy: $(F - \sigma_{F}) > (BL + \sigma_{BL})$. Finally, an additional delay of 0.5 days is added before and after the flare in order to take into 
account that the precise time of the flare is not known (1-day binned light curve). With this definition, a flare has a width of at least two days. With the hypothesis that the neutrino 
emission follows the gamma-ray emission, the signal time PDF is simply the normalized light curve of only the high state periods.

\begin{figure}[ht!]
\centering
\includegraphics[width=0.4\textwidth]{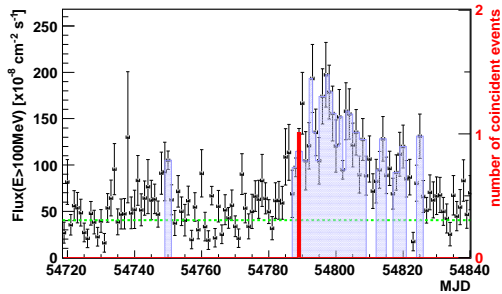}
\caption{Gamma-ray light curve (dots) of the blazar 3C279 measured by the LAT instrument onboard the Fermi satellite above 100 MeV. The light shaded histogram (blue)
indicated the high state periods. The dashed line (green) corresponds to the fitted baseline. The red histogram displays the time of the associated ANTARES neutrino event.
}
\label{fig:Result_3C279}
\end{figure}

For nine sources, no coincidences are found. 
For 3C279, a single high-energy neutrino event is found in coincidence. This event is located at 0.56$^{\circ}$ from the source location during a large flare in November 2008. 
The pre-trial p-value is 1.0~\%. Figure~\ref{fig:Result_3C279} shows the time distribution of the Fermi gamma-ray light curve of 3C279 and the time of the coincident neutrino event. 
This event was reconstructed with 89 hits distributed on ten lines with a track fit quality $\Lambda=-4.4$. The post-trial 
probability computed taking into account the ten searches is 10~\% and is thus compatible with background fluctuations.

\section{Summary}

This paper presents the first time-dependent search for cosmic neutrinos using the data taken with the full twelve line ANTARES detector during the last four months of 2008. 
For variable sources, time-dependent point searches are much more sensitive than time-integrated searches for variable sources due to the large reduction of the background. This 
search was applied to ten very bright and variable Fermi LAT blazars. The most significant correlation 
was found with a flare of 3C279 for which one neutrino event was detected in time/direction coincidence with the gamma-ray emission. 
The post-trial probability is about $10~\%$. Upper limits were obtained on the neutrino fluence for the ten selected sources.

%\vspace{\baselineskip}
\section{Acknowledgments}
I greatfully acknowledge the financial support of MICINN (FPA2009-13983-C02-01 and MultiDark 
CSD2009-00064) and of Generalitat Valenciana (Prometeo/2009/026).

\end{document}